# Fabrication of single-walled carbon nanotube/Si heterojunction solar cells with high photovoltaic conversion efficiency and stability

Feijiu Wang[1], Daichi Kozawa[1], Yuhei Miyauchi[1,2], Kazushi Hiraoka[3], Shinichiro Mouri[1], Yutaka Ohno[4], and Kazunari Matsuda[1*]

[1]*Institute of Advanced Energy, Kyoto University, Uji, Kyoto 611-0011, Japan*
[2]*Japan Science and Technology Agency, PRESTO, 4-1-8 Honcho Kawaguchi, Saitama 332-0012, Japan*
[3]*Hitachi Zosen Corporation, 2-11, Funamachi 2-Chome, Taisho-ku, Osaka 551-0022, Japan*
[4]*Department of Quantum Engineering, Nagoya University, Chikusa-ku, Nagoya 464-8603, Japan*

## ABSTRACT

The photovoltaic properties of carbon nanotube/Si heterojunction solar cells were investigated using network films of high quality single-walled carbon nanotubes (SWNTs) grown by atmospheric-pressure floating-catalyst chemical vapor deposition. Because of the optimization of the device window size and the utilization of SWNT thin films with both low resistivity and high transparency, a high photovoltaic conversion efficiency of greater than 12% was achieved for SWNTs/Si heterojunction solar cells without any post processing, such as carrier doping treatment. In addition, the high stability and reproducibility of the photovoltaic performance of these devices in air was demonstrated.

Carbon nanotubes have attracted a great deal of interest for photovoltaic applications because of their excellent optical and electronic properties, including the ability to tune their band gaps over a wide range by controlling the tube diameter [1] and their high carrier mobilities along their one-dimensional axes. [2-4] Solar cells based on simple heterostructures of single-walled carbon nanotubes (SWNTs) and Si have been extensively studied as model devices. [5-9] In the past few years, the photovoltaic performance of SWNTs/Si heterojunction solar cells has been improved using postprocessing techniques and complicated device structures, such as carrier doping of SWNT films by the infiltration of acid, [10] controlling the electronic junctions using an ionic liquid electrolyte, [11,12] and coating the Si surface with an anti-reflection layer. [13] Photovoltaic conversion efficiencies of greater than 10% have been reported for these devices. However, the need to use complicated postprocessing techniques is impracticable for the application of SWNTs to large-scale solar cells. In addition, such techniques may lead to less stable devices. Hence, the development of a method for the production of high efficiency, stable SWNTs/Si heterojunction solar cells that do not require postprocessing is strongly desired.

The physical properties of SWNTs thin films strongly affect the photovoltaic performance of SWNTs/Si heterostructure solar cells. It has been revealed that the SWNT film serves as a semi-transparent carrier transport layer in these solar cell devices. [14] Hence, one of the major factors limiting the photovoltaic efficiency of previously reported SWNT solar cells may be related to the carrier transport properties of the films. It is thus expected that SWNTs/Si heterostructure solar cells with these films that have high transparency and low sheet resistivity will improve photovoltaic conversion efficiencies.

Here, we report the results of our investigation of the photovoltaic properties of SWNTs/Si heterojunction solar cells using SWNT thin films fabricated using atmospheric-pressure floating-catalyst chemical vapor deposition (FC-CVD). A photovoltaic conversion efficiency of ~10% was obtained for solar cells using these high-quality SWNT thin films with both high transparency and low sheet resistance. In addition, various cells with SWNT films of different active window sizes and thicknesses were examined to determine the effect of these characteristics on the transparency of these films. Because of the optimization of such device parameters and the utilization of a high-quality SWNT film with both low resistivity and high transparency, a high photovoltaic conversion efficiency of greater than 12% was achieved in a SWNTs/Si heterojunction solar cell without any additional postprocessing, such as carrier doping. In addition, the high stability and reproducibility of the photovoltaic performance of this device in air was demonstrated.

SWNTs were grown using the FC-CVD method, as described previously.[15,16] Carbon monoxide (CO) was the carbon source, and the catalyst particles were produced by the decomposition of ferrocene vapor. CO (100 sccm) was passed through a temperature-controllable cartridge containing ferrocene powder. Additional CO (300 sccm) and $CO_2$ (3 sccm) were also introduced to the furnace, and the growth temperature was set at 850 °C. The SWNTs film was collected at room temperature on a membrane filter comprising cellulose acetate and nitrocellulose. The thickness of a SWNT film can be controlled by the collection time.

An *n*-Si substrate (1–10 Ω·cm) was used for the solar cell devices. The native $SiO_2$ layer of the active window was removed using a buffered HF solution. The membrane filter containing the SWNT film was then placed on the Si substrate such that the film was in direct contact with the top surface of the substrate. A fiber wiper was then applied to the surface of the substrate with adequate pressure to release the film from the membrane filter and to create a connection between the film and the substrate. It was essential to place a drop of ethanol on the surface to ensure good contact between the films and the substrates and to densify the SWNT film. [17] SWNT films with window sizes ($\phi$: diameter) ranging from 1 to 3 mm were used. Next, each SWNT film was connected to gold (Au) as the anode, while each *n*-Si substrate was connected to indium (In) as the cathode.

The optical transmission spectra of the SWNT films on the glass substrates were measured using a UV/Vis/NIR spectrophotometer (HITACHI U-4100). To evaluate the photovoltaic properties of the devices, the SWNTs/Si heterojunction solar cells were irradiated using a solar simulator (San-Ei Electric XES-40S1) under AM 1.5 conditions (~100 mW/cm$^2$), and the current density–voltage (*J–V*) data were recorded using a sourcemeter (Keithley 2400). The AM 1.5 (100 mW/cm$^2$) conditions in the solar simulator were confirmed using a standard cell (BS-500BK). The sheet resistance of the SWNT network was measured using a digital multimeter (Keithley 2100) connected to a four-probe collinear arrangement (Astellatech SR4-S).

Figure 1(a) shows the schematic of the device structure of a SWNTs/Si heterojunction solar cell and (b) exhibits from cross−section view. Figure 1(c) shows photographs of uniform and large area (12.56 cm$^2$) SWNT films fabricated at collection times of 12, 16, 20, 30, 40, and 60 min. The corresponding transmittance spectra of these films are shown in the inset of Fig. 1(d). The transmission value for each film at 550 nm (*T*) is indicated in the inset of Fig. 1(d). Figure 1(d) shows the sheet resistances of the SWNT films as a function of *T*. The SWNT films fabricated by the FC-CVD method exhibit both low resistivity and high transparency, such as 2 kΩ/□ for *T* = 80%, which is significantly superior to those of films prepared using a spray method. [18]

The effect of the thickness of the SWNT film on the photovoltaic properties of SWNTs/Si heterojunction solar cells was investigated. Figure 2(a) shows the *J−V* curves of SWNTs/Si heterojunction solar cells with an active window size (*ϕ*) of 2 mm fabricated using SWNT films with different thicknesses (transmission values *T*). The *J−V* curves for the SWNTs/Si heterojunction solar cells drastically change as the transmittance of the SWNT film varies. Based on the evaluation of this series of *J−V* curves, it is found that the photovoltaic conversion efficiency *η* strongly depends on the thickness of the film (i.e., the transmittance *T*). Figure 2(b) shows the values obtained for *η*, the fill factor (FF), the short circuit current density ($J_{sc}$), and the open circuit voltage ($V_{oc}$) as a function of *T* derived from this series of *J−V* curves. The photovoltaic conversion efficiency *η* increases as *T* decrease and reaches a maximum value at *T* ~ 80%. In contrast, a negligible change in $V_{oc}$ is observed, while the FF drastically increases as *T* decreases. In addition, $J_{sc}$ drastically increases as *T* decreases in the high-transmittance region and moderately decreases as *T* decreases in the low-transmittance region. Thus, increase in both FF and $J_{sc}$ from 26.5% to 71% and 18.9 to 26.9 mA/cm$^2$, respectively, contributes to the drastic increase in *η* from 2.5% to 10.6% from $T$ = 97% to 80%. It should be noted that the photovoltaic conversion efficiency of the cell with a high transparency and low resistivity SWNT film prepared using FC-CVD ($T$ = 80%) reached a value of 10.6%, which is much larger than the value of 3.2% obtained for cells prepared using a spray method ($T$ ~80%). [18]

Next, the change in the conversion efficiency of solar cells as a function of the window size and SWNT film transmittance was investigated. The SWNTs/Si heterojunction solar cells with active window sizes ($\phi$) ranging from 1 to 3 mm and SWNT films with transmittance ($T$) ranging from 70% to 97% were prepared (Fig. S1-S3). The photovoltaic conversion efficiencies of solar cells with SWNT films of different transmittance are shown in Fig. 3 as a function of the window size. The figure shows that the photovoltaic conversion efficiency increases as the active window size decreases, mainly because of a reduction in the effective series resistance, as discussed below. However, the optimized $T$ values are different depending on the active window size. Thus, the optimized device parameters for SWNTs/Si heterojunction solar cells ($\phi$ = 1 mm and $T$ = 91%) are determined based on these results and are shown to provide a high conversion efficiency of 12.2%.

To gain further insight into the impact of the properties of the SWNT film on the photovoltaic performance of SWNTs/Si heterojunction solar cells, the series resistance ($R_s$) of the optimized devices ($\phi$ = 1 mm) fabricated by different thickness SWNT films was calculated using the $J$–$V$ curve data and the following relationship: [20,23]

$$I \frac{dV}{dI} = R_S I. \tag{1}$$

Figure 4(a) shows that the series resistance increases as $T$ increases (decreased thickness of the SWNT network film). This result suggests that the decrease in the FF with an increase in $T$ (Fig. 2(b)) is because of the increase in $R_s$. The plot of $I(dV/dI)$ for the device fabricated using a SWNT film with $T = 91\%$ as a function of $I$ is shown in the inset of Fig. 4(a). The value for $R_s$ (= 180 Ω) corresponding to 1.41 Ω·cm$^2$ is obtained from the slope of the line in the inset of Fig. 4. This low series resistance is mainly because of the high density and close connection of SWNTs network, which occurs because of Y-junction formation in the films fabricated using the FC-CVD method. [16]

The diode ideality factor ($n$) was also evaluated to investigate the carrier transport mechanism of the devices. This factor is determined under forward bias conditions using the relationship described below, [19]

$$\frac{1}{n} = \frac{kT}{q} \cdot \frac{d \ln I_{dark}}{dV}, \quad (2)$$

where $q$ is the electronic charge, $k$ is the Boltzmann constant, $V$ is the applied voltage, and $I_{dark}$ is the dark current. The plot of $\ln I_{dark}$ for the device fabricated using the film with $T = 91\%$ as a function of $V$ is shown in the inset of Fig. 4(b). From the slope of this line, the diode ideality factor $n$ is determined to be 1.65. Figure 4(b) shows the values for $n$ obtained for the devices as a function of $T$. The figure shows that $n$ moderately decreases from 1.93 to 1.36 as $T$ decreases (i.e., the thickness of the SWNT film increased). These values are much smaller than those previously reported, [18,20] and nearly the same as those of typical Si solar cells. [21,22] Based on these results, it can be concluded that $n$ is related to the diffusion recombination process, accompanied by tunneling. [23] The moderate increase in $n$ with increasing $T$ is because of the increase in the sheet resistance of these films. [24]

Figure 5(a) shows a typical $J-V$ curve of the optimized SWNTs/Si heterojunction solar cell ($\phi = 1$ mm) using the SWNT film with $T = 91\%$ under AM 1.5 conditions. The values for $J_{sc}$, $V_{oc}$, and FF extracted from the $J-V$ curve are 30.2 mA/cm$^2$, 0.586 V, and 70%, respectively, which result in a high photovoltaic conversion efficiency $\eta$ of 12.5%. This value (12.5%) is higher than the previously reported value (10.5 %), [11] and is the highest photovoltaic conversion efficiency for a SWNT/Si heterojunction solar cell fabricated using as-grown SWNTs without any carrier doping treatment as a postprocessing technique. Previous studies of the photovoltaic conversion process have revealed that the SWNT film mainly serves as a semi-transparent carrier transport layer in SWNT/Si heterojunction solar cells. [25] Thus, the high photovoltaic conversion efficiency of greater than 12% is because of the high quality (low resistivity and high transparency) of the SWNT film.

The photovoltaic conversion efficiency of 20 solar cells fabricated using the optimized device parameters was then measured, and a remarkable reproducibility for $\eta$ in the narrow range from 11.8% to 12.7% is observed, as shown in the inset of Fig. 5(a). The stability of the photovoltaic performance in the SWNT/Si heterojunction solar cell was also examined. Figure 5(b) shows the time evolution of the photovoltaic conversion efficiency $\eta$, $J_{sc}$ and $V_{oc}$ of the solar cell. The values for $J_{sc}$ and $V_{oc}$ nearly remain the same, while the FF moderately decreases with prolonged time, resulting in a modest decay in the photovoltaic conversion efficiency from 12.5% to 10.5% over 10 days. This decreased rate of 0.2% per day is much smaller than the previously reported value of ~1.2% per day for SWNTs/Si solar cells doped with HNO$_3$. Thus, the optimized

SWNTs/Si solar cells fabricated using SWNT films grown by FC-CVD that have low resistivity and high transparency show both superior photovoltaic performance (conversion efficiency > 12%) and high stability. Also, with $HNO_3$ doping, the optimized SWNTs/Si heterojunction solar cell exhibits a very high photovoltaic conversion efficiency of 14.5% (Fig. S4).

In summary, SWNTs/Si heterojunction solar cells prepared using high quality SWNT films with low resistivity and high transparency fabricated by the FC-CVD method were investigated. The optimum device structure was determined to consist of 1-mm-diameter window and a SWNT film with $T = 91\%$, and a photovoltaic conversion efficiency of 12% was obtained for a SWNTs/Si heterojunction solar cell prepared without postprocessing using this structure. In addition, the high reproducibility of the photovoltaic performance of this type of device and its stability in air were clearly demonstrated, suggesting the superior potential of carbon nanotube solar cells.

**ACKNOWLEDGMENTS**

YM was financially supported by PRESTO, JST. This study was supported by a Grant-in-Aid for Scientific Research from JSPS (Grant Nos. 22740195, 23340085, 24681031, and 40311435), Yazaki Memorial Foundation for Science and Technology, Nippon Sheet Glass Foundation for Materials Science and Engineering, and The Canon Foundation.

# REFERENCES

[1] R. Saito, G. Dresselhaus, and M. S. Dresselhaus, *Physical Properties of Carbon Nanotubes* (Imperial College, London, 1998).

[2] M. Kanungo, H. Lu, G. G. Malliaras, and G. B. Blanchet, Science **323**, 234 (2009).

[3] M. S. Fuhrer, B. M. Kim, T. Durkop, and T. Brintlinger, Nano Lett. **2**, 755 (2002).

[4] E. S. Snow, P. M. Campbell, and M. G. Ancona, Appl. Phys. Lett. **86**, 033105 (2005).

[5] P. L. Ong, W. B. Euler, and I. A. Levitsky, Nanotechnology **22**, 105203 (2010).

[6] D. Kozawa, K. Hiraoka, Y. Miyauchi, S. Mouri, and K. Matsuda, Appl. Phys. Express **5**, 042304 (2012).

[7] Y. Jung, X. K. Li, N. K. Rajan, A. Taylor, and M. A. Reed, Nano Lett. **13**, 95 (2013).

[8] Z. R. Li, V. P. Kunets, V. Saini, Y. Xu, E. Dervishi, G. J. Salamo, A. R. Biris, and A. S. Biris, ACS Nano **3**, 1407 (2009).

[9] D. D. Tune, B. S. Flavel, R. Krupke, and J. G. Shapter, Adv. Energy Mater. **2**, 1043 (2012).

[10] Y. Jia, A. Cao, X. Bai, Z. Li, L. Zhang, N. Guo, J. Wei, K. Wang, H. Zhu, D. Wu, and P. M. Ajayan, Nano Lett. **11,** 1901 (2011).

[11] P. Wadhwa, B. Liu, M. A. McCarthy, Z. Wu, and A. G. Rinzler, Nano Lett. **10**, 5001 (2010).

[12] P. Wadhwa, G. Seol, M. K. Petterson, J. Guo, and A. G. Rinzler, Nano Lett. **11**, 2419 (2011).

[13] E. Shi, L. Zhang, Z. Li, P. Li, Y. Y. Shang, Y. Jia, J. Wei, K. Wang, H. Zhu, D. Wu, S. Zhang, and A. Cao, Sci. Rep. **2**, 884 (2012).

[14] S. Jeong, E. C. Garnett, S. Wang, Z. Yu, S. Fan, M. L. Brongersma, M. D. McGehee, and Y. Cui, Nano Lett. **12**, 2971 (2012).

[15] A. Moisala, A. G. Nasibulin, D. P. Brown, H. Jiang, L. Khriachtchev, and E. I.

Kauppinen, Chem. Eng. Sci. **61**, 4393 (2006).

[16] D.-M. Sun, M. Y. Timmermans, Y. Tian, A. G. Nasibulin, E. I. Kauppinen, S. Kishimoto, T. Mizutani, and Y. Ohno, Nature Nanotech. **6**, 156 (2011).

[17] D.-M. Sun, M. Y. Timmermans, A. Kaskela, A. G. Nasibulin, S. Kishimoto, T. Mizutani, E. I. Kauppinen, and Y. Ohno, Nature Commun. **4**, 2302 (2013).

[18] F. Wang, D. Kozawa, Y. Miyauchi, K. Hiraoka, S. Mouri, and K. Matsuda, Appl. Phys. Express **6**, 102301 (2013).

[19] J. Nelson, *The physics of solar cells* (Imperial College, London, 2003).

[20] Y. Jia, J. Wei, K. Wang, A. Cao, Q. Shu, X. Gui, Y. Zhu, D. Zhuang, G. Zhang, B. Ma, L. Wang, W. Liu, Z. Wang, J. Luo, and D. Wu, Adv. Mater. **20**, 4594 (2008).

[21] M. A. Green, A. W. Blakers, J. Shi, E. M. Keller, and S. R. Wenham, Appl. Phys. Lett. **44**, 1163 (1984).

[22] E. Radziemska, Energy Convers. Manage. **46**, 1485 (2005).

[23] A. E. Rakhshani, J. Appl. Phys. **90**, 4265 (2001).

[24] C. Xie, X. Zhang, Y. Wu, X. Zhang, X. Zhang, Y. Wang, W. Zhang, P. Gao, Y. Han, and J. Jie, J. Mater. Chem. A **1**, 8567 (2013).

[25] D. D. Tune, F. Hennrich, S. Dehm, M. F. G. Klein, K. Glaser, A. Colsmann, J. G. Shapter, U. Lemmer, M. M. Kappes, R. Krupke, and B. S. Flavel, Adv. Energy Mater. **3**, 1091 (2013).

[26] See supplementary material at [*URL will be inserted by AIP*] for *variations of window size dates and HNO$_3$ doped date.*

**FIGURE CAPTIONS**

**Fig. 1.** (a) Schematic of the device structure of a SWNTs/Si heterojunction solar cell. (b) Cross-sectional view of the device structure. (c) Photographs of SWNT films with transmittance values $T$ = 97, 94, 91, 86, 80, and 70% at 550 nm, respectively. (d) Sheet resistance of SWNT films as a function of $T$. Inset shows the transmission spectra of the SWNT films.

**Fig. 2.** (a) Current density–voltage ($J$–$V$) curves for heterojunction solar cells fabricated using SWNT films of various thicknesses (transmittance). The conversion efficiency of the solar cells and the corresponding transmittance values for the SWNT films are shown in the figure. (b) Values for $\eta$, FF, $J_{sc}$, and $V_{oc}$ obtained from the $J$–$V$ curves for each device as a function of $T$.

**Fig. 3.** Photovoltaic conversion efficiency of solar cells with SWNT films of various thicknesses as a function of the window size.

**Fig. 4.** (a) Series resistance of the devices plotted as a function of the transmittance $T$. The inset shows $I(dV/dI)$ as a function of $I$ for the device fabricated using the SWNT film with $T$ = 91%. The linear fitted result is shown as a solid line, and the slope of the fitted line corresponds to the cell series resistance. (b) Diode ideality factors as a function of the transmittance $T$. The inset shows the semi-logarithmic plot of the $J$−$V$ curve for the device fabricated using the SWNT film with $T$ = 91%. The linear fitted line is shown as the solid line.

**Fig. 5.** (a) Current density–voltage (*J–V*) curves for optimized SWNTs/Si heterojunction solar cells fabricated with a windows size of 1 mm using a SWNT film with $T$ = 91%. The inset shows a histogram of the variation in the photovoltaic conversion efficiency of the fabricated SWNTs/Si solar cell devices. (b) Time evolution of the photovoltaic properties FF, $V_{OC}$, $J_{SC}$, and $\eta$ of an optimally prepared device.

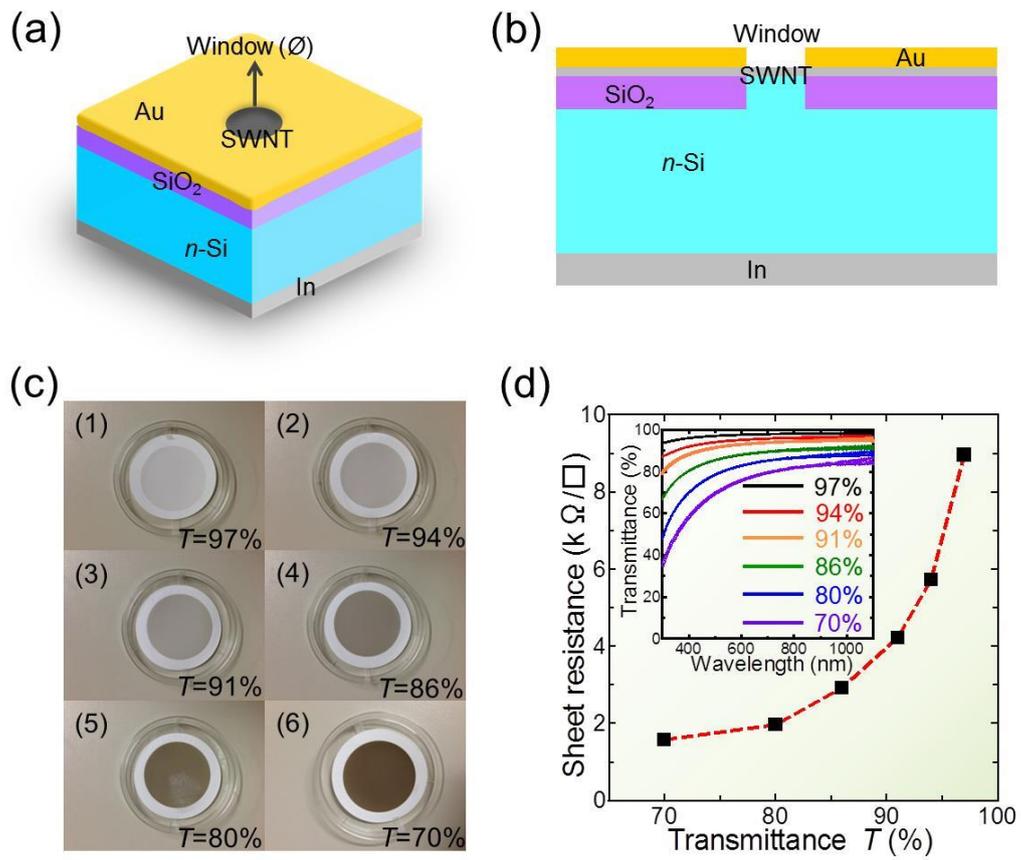

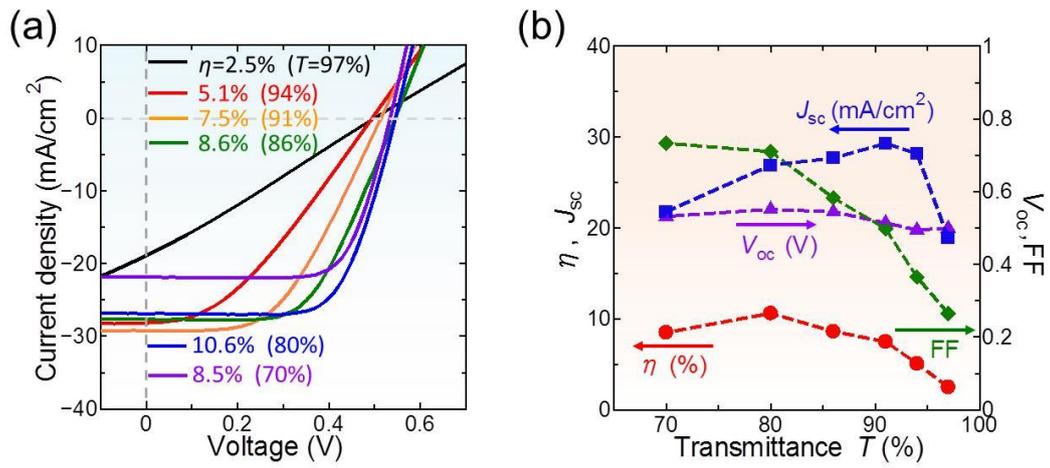

Fig. 2  F. Wang et al.

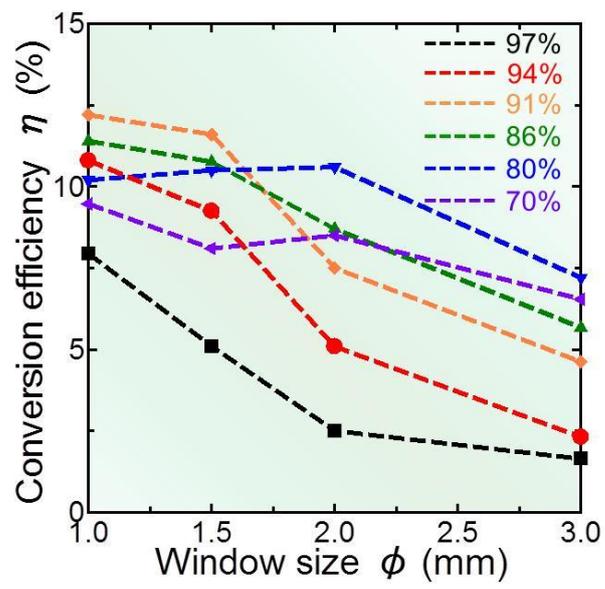

Fig.4 F. Wang et al.,

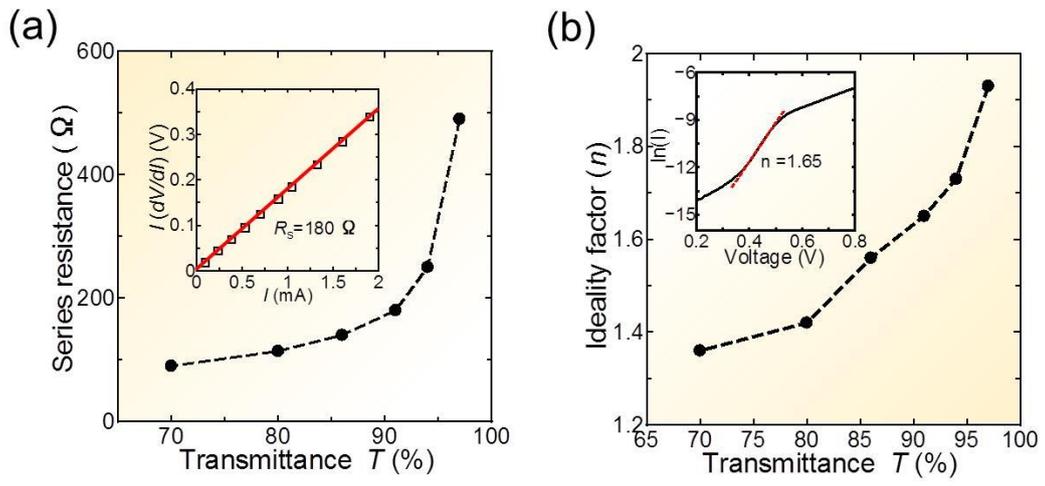

Fig. 4  F. Wang et al.,

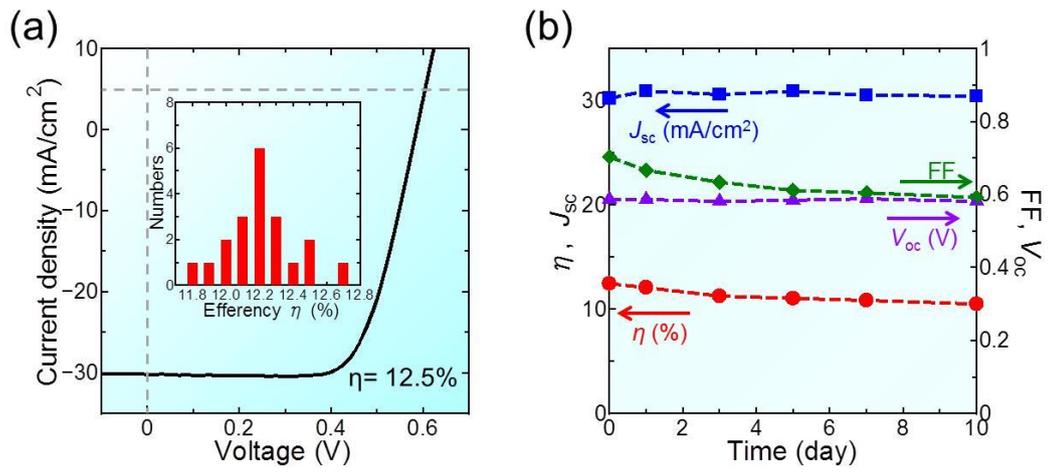

Fig. 5  F. Wang et al.

# Supplementary

**Fabrication of single-walled carbon nanotube/Si heterojunction solar cell with high photovoltaic conversion efficiency and stability**

Feijiu Wang[1], Daichi Kozawa[1], Yuhei Miyauchi[1,2], Kazushi Hiraoka[3], Shinichiro Mouri[1], Yutaka Ohno[4], and Kazunari Matsuda[1,2*]

[1]*Institute of Advanced Energy, Kyoto University, Uji, Kyoto 611-0011, Japan*
[2]*Japan Science and Technology Agency, PRESTO, 4-1-8 Honcho Kawaguchi, Saitama 332-0012, Japan*
[3]*Hitachi Zosen Corporation, 2-11, Funamachi 2-Chome, Taisho-ku, Osaka 551-0022, Japan*
[4]*Department of Quantum Engineering, Nagoya University, Chikusa-ku, Nagoya 464-8603, Japan*

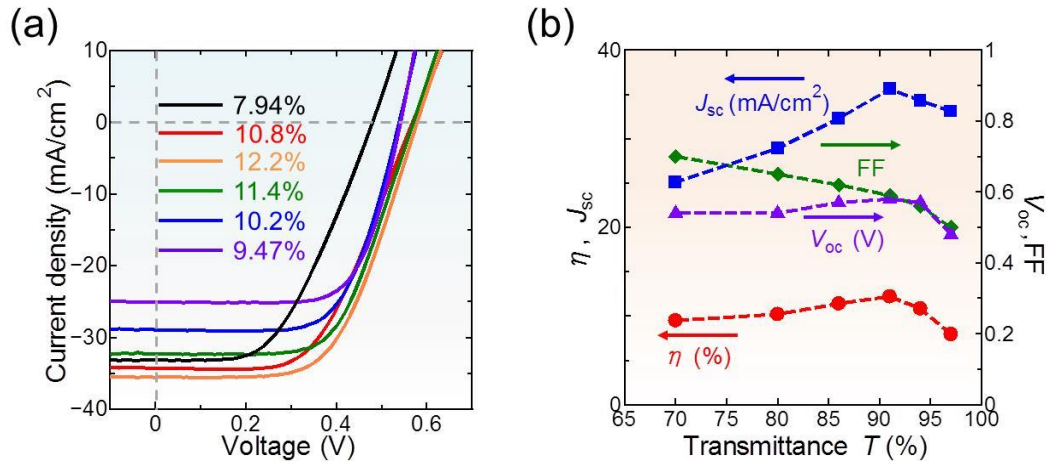

**Fig. S1** (a) Current density–voltage (*J*–*V*) curves for solar cells with 1-mm-diameter active windows fabricated using SWNT films of varying thicknesses (transmittance). (b) Values for $\eta$, FF, $J_{sc}$, and $V_{oc}$ obtained from the *J*–*V* curves for each device as a function of *T*.

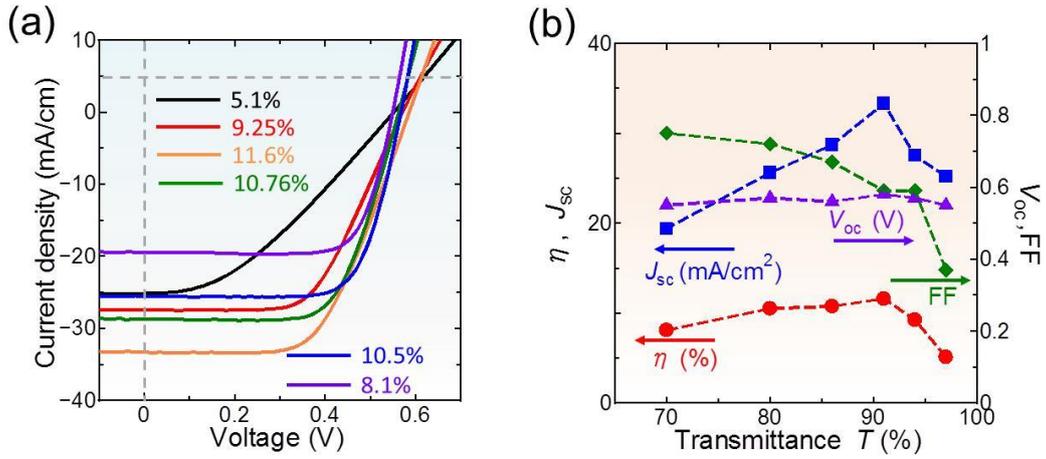

**Fig. S2** (a) Current density–voltage ($J$–$V$) curves for heterojunction solar cells with 1.5-mm-diameter active windows fabricated using SWNT films of varying thickness (transmittance). (b) Values for $\eta$, FF, $J_{sc}$, and $V_{oc}$ obtained from the $J$–$V$ curves for each device as a function of $T$.

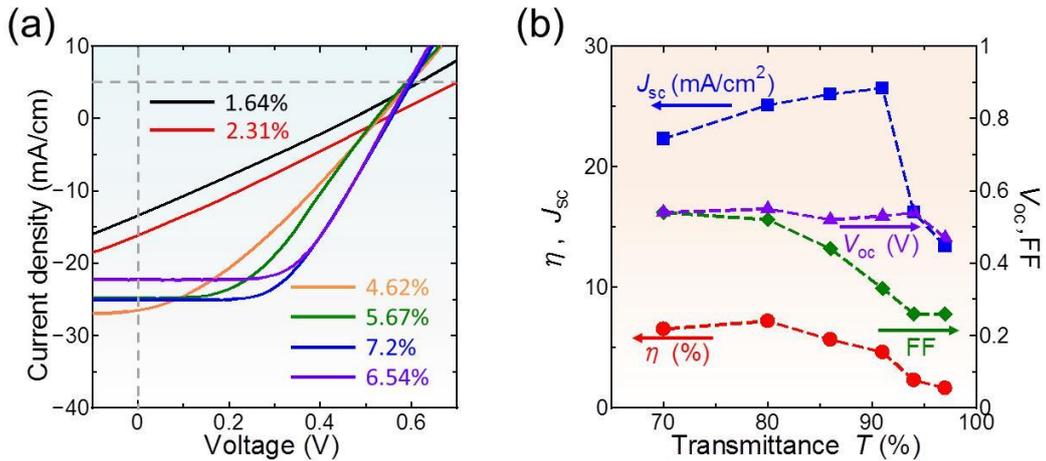

**Fig. S3.** (a) Current density–voltage (J–$V$) curves for solar cells with 3-mm-diameter active windows fabricated using networked SWNT films of varying thickness (transmittance). (b) Values for $\eta$, FF, $J_{sc}$, and $V_{oc}$ obtained from the $J$–$V$ curves for each device as a function of $T$.

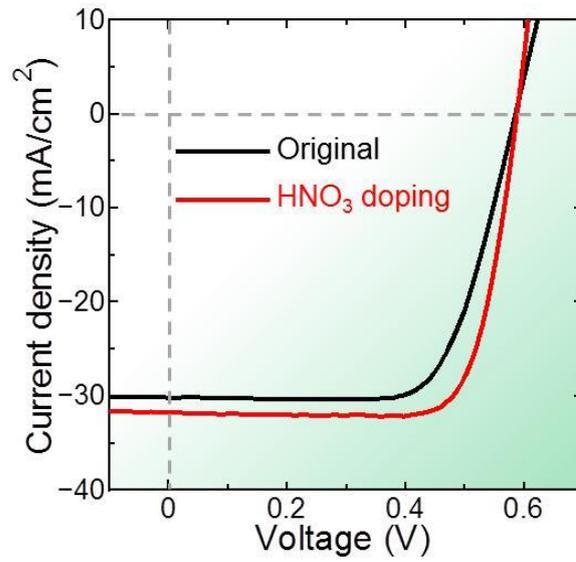

**Fig. S4.** Current density–voltage (*J*–*V*) curves for original (black line) and HNO$_3$-doped solar cells with 1-mm-diameter active windows fabricated using networked SWNT films with a transmittance of 91%.